\documentclass[twocolumn,showpacs,preprintnumbers,amsmath,amssymb]{revtex4}
\usepackage{graphicx}
\usepackage{bm}
\usepackage{epsf}
\usepackage{xcolor}
\usepackage{epstopdf}
\begin{document}
\title{The influence of Gaussian pinning on the melting scenario of a two-dimensional
soft-disk system: First-Order versus Continuous Transition}

\author{Yu. D. Fomin \footnote{Corresponding author: fomin314@mail.ru}}
\affiliation{The Vereshchagin Institute of High Pressure Physics,
Russian Academy of Sciences, Kaluzhskoe shosse, 14, Troitsk,
Moscow, 108840, Russia }

\author{Eu. A. Gaiduk }
\affiliation{The Vereshchagin Institute of High Pressure Physics,
Russian Academy of Sciences, Kaluzhskoe shosse, 14, Troitsk,
Moscow, 108840, Russia }

\author{E. N. Tsiok}
\affiliation{The Vereshchagin Institute of High Pressure Physics,
Russian Academy of Sciences, Kaluzhskoe shosse, 14, Troitsk,
Moscow, 108840, Russia }

\author{V. N. Ryzhov }
\affiliation{The Vereshchagin Institute of High Pressure Physics,
Russian Academy of Sciences, Kaluzhskoe shosse, 14, Troitsk,
Moscow, 108840, Russia }
\date{\today}

\begin{abstract}
Two-dimensional systems are realized experimentally as thin layers
on a substrate. The substrate can have some imperfections (defects
of the crystalline structure, chemical impurities, etc.), which
demonstrate stronger interaction with the particles of the
two-dimensional layer than the rest of the system. Such randomly
distributed centers of strong interactions are called "pinning
centers". The presence of random pinning can substantially change
the behavior of the system. It not only shifts the melting point
of the system, but can also change the melting
scenario itself. In the present paper the influence of Gaussian
pinning on the melting scenario of a two-dimensional system of
soft disks is studied by means of molecular dynamics simulation.
We randomly introduce into the system of soft disks a set of
"pinning centers" which attract the particles via the Gauss
potential. We observe that increasing the depth of a Gaussian
well leads to a change in the melting scenario of the system. The results demonstrate that simple kind of quenched disorder can significantly affect the melting scenario of two-dimensional systems, offering the possibility of its introduction in complex experiments and studying its influence on the self-assembly and phase diagram of two-dimensional systems in rotating external fields.

{\bf Keywords}: Molecular dynamics simulations; Two-dimensional system; Phase diagram; Phase transitions; Two-dimensional melting, Gaussian pinning.

\end{abstract}

\pacs{61.20.Gy, 61.20.Ne, 64.60.Kw}

\maketitle


\section{Introduction}

Pinning is a powerful tool that can significantly affect not only
the melting and crystallization scenarios of two-dimensional (2D) systems
but also the phase diagram. Due to strongly developed fluctuations
and the presence of coupled topological defects, dislocation
pairs, 2D crystals are already endowed with quasi-long-range
translational order, the correlation functions of which decay
algebraically. However, the order in bonds between a particle and
its nearest neighbors, that is, the orientational order, is
preserved at a long range. As the temperature increases, the
dissociation of coupled dislocation pairs occurs, which leads to
the complete destruction of translational order while preserving the
quasi-long-range of orientational order. In this case, the crystal
passes continuously into orientationally ordered liquid with a
zero shear modulus, which is called a hexatic phase. A further
increase in temperature leads to the destruction of orientational
order due to the dissociation of bounded pairs of disclinations,
and the continuous transition of the hexatic phase into
isotropic liquid takes place. A cascade of two continuous
transitions through an intermediate hexatic phase formed the basis
of the classical
Berezinskii-Kosterlitz-Thouless-Halperin-Nelson-Young (BKTHNY)
theory of 2D melting \cite{berez, koster, halpnel1, halpnel2,
young}.

As in the case of three dimensions, 2D crystals can melt into
isotropic liquid through a first-order transition, for example, as
a result of the dissociation of disclination quadrupoles or the
formation of grain boundaries \cite{chui,ryzhov1,ryzhov2}.

Later, Bernard and Krauth (BK) presented a study of
melting of a 2D hard-disk system and found a new two-stage
melting scenario, according to which a first-order phase
transition occurs between isotropic liquid and the hexatic
phase, and the crystal-to-hexatic phase transition occurs via a
continuous Berezinskii-Kosterlitz-Thouless (BKT) transition
\cite{bernard,engel,kapfer,ourphysusp,ourjetp}. A consistent microscopic explanation of
the mechanism of the scenario is currently lacking.

It should be noted that in the conditions of a real experiment, in
contrast to most theoretical and computer studies, the presence of
certain random factors (substrate defects, impurities, field
inhomogeneities, etc.) interacting with the main particles of the
system is likely to take place in the system. Attempts to theoretically describe
the effect of such a random disorder in 2D systems have been made
since the emergence of theories of 2D melting \cite{nelson1,nelson2,cha}.
By modifying the organization of pinning, it is possible to
radically change the melting scenario of a 2D system
\cite{ourphysusp,ourjetp,tsiokpre1,tsiokpre2}. With regard to pinned particles at random sites,
including interstitial lattice sites, it was shown that the BKTHNY
melting scenario persisted and that the solid phase was destroyed
entirely for high pinning fractions (see \cite{nelson1, nelson2,
cha,softmat}). Experiments and simulations of 2D melting of
super-paramagnetic colloidal particles with quenched disorder
confirmed the increased stability range of the hexatic phase (see
\cite{maretprl, maretpre}). Also, computer modeling
of the behavior of systems with different repulsive potentials in
the presence of random pinning showed that pinning practically did
not affect the stability limit of the hexatic phase with respect
to isotropic liquid but lowered the limit of stability of the
crystal due to the destruction of the quasi-long-range
translational order \cite{tsiokpre1, tsiokpre2}. The influence of
random pinning on 2D melting via first-order phase transition
was investigated in 2D systems with core-softened potential
\cite{tsiokpre1} and the Hertz potential with $\alpha = 5/2$,
which describes 2D colloidal systems with elastic repulsion
between particles \cite{tsiokpre2}. It was demonstrated that
random pinning could induce the hexatic phase in the solid phase and transform
first-order melting into the BK scenario. An even more interesting
result was obtained in \cite{tsiokpre2} where it was shown that
random pinning significantly changed the transition mechanisms
between two crystalline phases at very low temperatures. Whereas
in a system without random pinning, the transition occurs as a
first-order transition, the introduction of pinning makes this
transition a three-stage one: the triangular crystal continuously
transforms into a hexatic, the hexatic transforms into a tetratic
through a first-order transition, after which the tetratic
continuously transforms into a square crystal.

The opposite case was considered in \cite{dijkstra} where the
melting of a 2D system of hard disks with quenched disorder, which
results in pinning random particles on the crystal lattice, was
studied. This kind of pinning stabilizes the solid phase and can
destroy the hexatic phase. We are not aware of real
experiments with this kind of disorder.

An important circumstance in works on computer simulation
\cite{maretprl, tsiokpre1, tsiokpre2, dijkstra} is the choice of a
model for random inhomogeneities interacting with system
particles. As a rule, the model of disorder was delta function-like wells, each
of which contained one particle. A particle falling into such a
well remains motionless. Pinned particles behave in such system
like particles with zero temperature. Such a model is a kind of a
limiting case. In fact, the strength of interaction with
disorder is finite, and pinning is not strict. In addition,
several particles can be drawn into the area of action of this
force. To describe the interaction of such pinning with the
particles of the system we used randomly located attractive wells
of finite depth, having a Gaussian shape (similar to the
experiment).

Earlier, within the framework of computer simulation by the method
of molecular dynamics, we studied a 2D soft disk system with the potential $U(r)=\varepsilon \left(
\frac{\sigma}{r} \right)^{12}; n =
12$ with and without random pinning \cite{n12}. As a result, it
was shown that a 2D system of soft disks melted
according to the BK melting scenario
at $T = 1.0$. Random pinning affected the density of crystal
destruction, shifting the existence region of the crystal to
higher densities along the $T = 1.0$ isotherm, but the melting
scenario did not change. In the present work, we study the
melting of 2D soft disk systems with $n = 12$ in the presence of
Gaussian pinning depending on the depth of the well by classical
molecular dynamics. Comparative analysis of the influence of
Gaussian pinning and delta function-like pinning on the phase diagram and the
melting scenario of 2D soft disks with $n = 12$ is carried out.

\section{System and Methods}

In the present work we simulated a system of 20000 soft disks
interacting via the potential $U(r)=\varepsilon \left(
\frac{\sigma}{r} \right)^{12}$ in a rectangular box with a periodic
boundary condition. Parameters $\varepsilon$ and $\sigma$ can
be used as energy and length scales respectively,
a cutoff $r_c/\sigma$ = 5.0. Below we used
reduced units based on these parameters. The initial structure of
the system is a perfect triangular crystal.

\begin{figure}
\includegraphics[width=10cm]{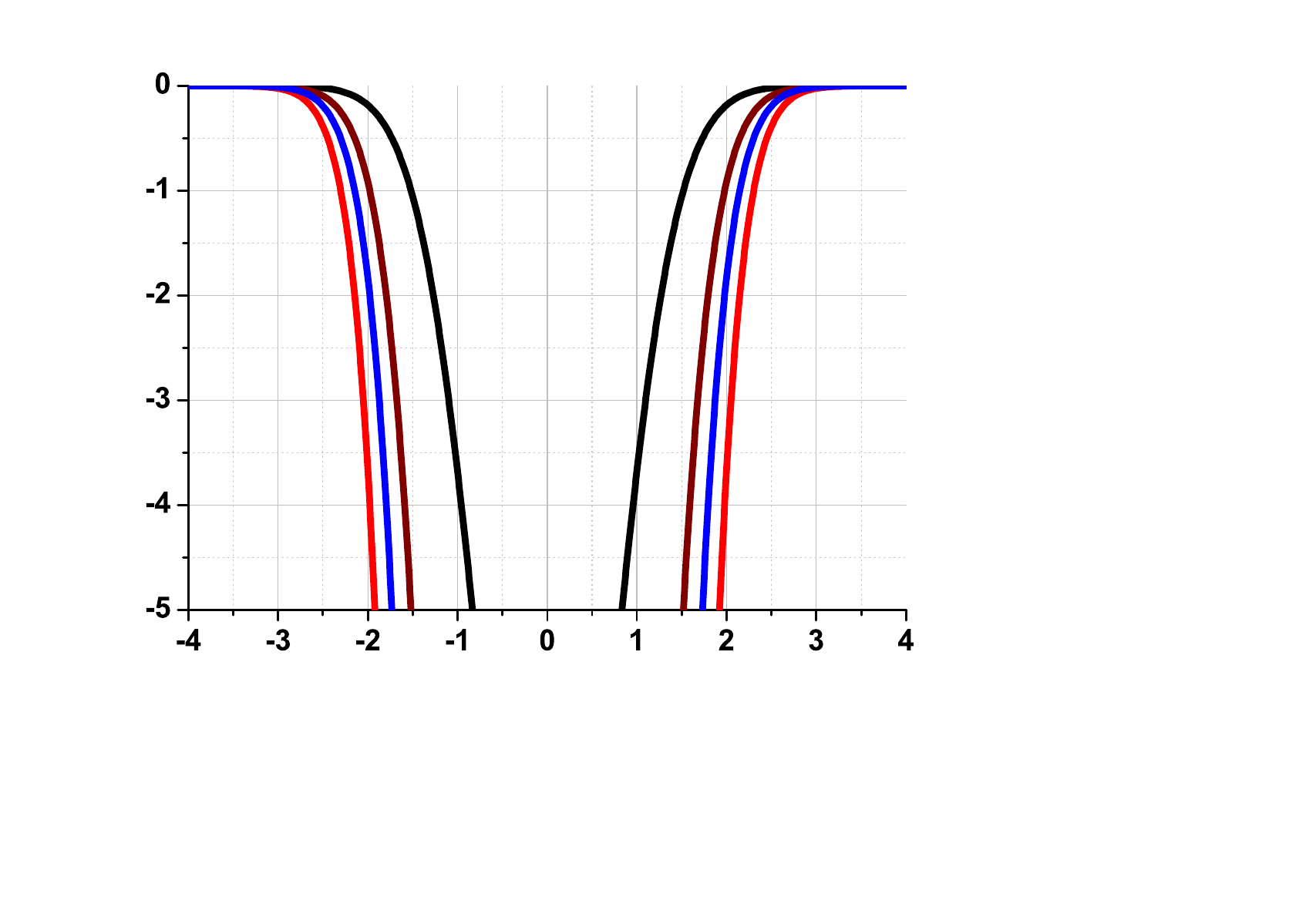}%

\caption{\label{gp} The Gauss potential depending on the well
depth $w$. The black line is $w=10.0$.The wine line is $w=50.0$.
The blue line is $w=100.0$. The red line is $w=200.0$. }
\end{figure}

We introduced 20 pinning centers ($0.1 \%$ of the system size) into
the system. The coordinates of these pinning centers are taken
randomly. The pinning centers interact with the particles of the
system via the Gauss potential:

\begin{equation}
U_{pin}=-w \cdot e^{-r^2},
\end{equation}
where $w$ is the depth of the well and $r$ is the distance between
the pinning center and a particle of the system in the reduced units $r/\sigma$.
We considered cases with $w=1.0$, $10.0$, $50.0$,
$100.0$, and $200.0$. Due to attraction to the pinning centers, more dense ranges (clusters) are formed in the vicinity of these centers. In fact, it is not so much the depth of the well as the width of the well, that is, the radius of action of the Gaussian force,
$r_{cg}$, that determines the size and structure of the cluster near
the pinning center. All simulations are
performed at constant temperature $T=1.0$. Therefore, it can be
assumed that even at $U_{pin}=-2.0$, the particles that fall within
$r_{cg}$ will be in the cluster throughout the entire simulation
time at $T=1.0$. So, for $U_{pin}=-2.0$, the value of $r_{cg}=1.27$
for $w=10.0$, $r_{cg}=1.8$ for $w=50.0$, $r_{cg}=1.96$ for $w=100.0$,
and $r_{cg}=2.14$ for $w=200.0$. Since the values of $r_{cg}$ for
$w=50.0$, $100.0$, and $200.0$ differ slightly, it can be
predicted that the size and structure of the clusters will be
similar. The Gauss potential for different $w$ is shown in Fig.
\ref{gp}.

The systems are simulated in the canonical ensemble (constant number of particles N, volume V and temperature T) with a Nose-Hoover thermostat for $2 \cdot 10^7$ steps for
equilibration and more $3 \cdot 10^7$ steps for calculation of its
properties. The timestep is set to $dt=0.001$. Firstly, we simulate
the system in a wide range of densities with a large step in order
to roughly identify the melting region. After that the region of
interest is simulated with a small step $\Delta \rho=0.002$. We
calculate the equation of state, i.e. the dependence of pressure
on the density, the translational and orientational correlation
functions defined below, which are used to identify the melting
scenario in the system under consideration.

As it was shown in our recent work \cite{soft-av}, correct
determination of the melting point of a 2D system requires large
statistical averaging. For this reason, all simulations were
performed ten times with different initial conditions: different
location of the pinning centers and different initial velocities
of the particles. All reported results are obtained by averaging
over these 10 replicas.

As it was mentioned above, 2D crystals demonstrate long-range
orientational and quasi-long-range translational order. These two
orders are characterized by corresponding correlation functions.

The translational correlation function is defined as

\begin{equation}
g _T(r)=\left< \frac{<\exp(i{\bf G}({\bf r}_i-{\bf r}_j))>}{g(r)} \right>_{replica},
\label{GT}
\end{equation}
where $r=|{\bf r}_i-{\bf r}_j|$, {\bf G} is a reciprocal lattice vector and $g(r)$ is the radial distribution function of the system. In the crystalline phase
the translational correlation function demonstrates algebraic decay: $g_T(r)\propto r^{-\eta_T}$ with
$\eta_T \leq \frac{1}{3}$ \cite{halpnel1, halpnel2}, while in the hexatic phase and isotropic liquid $g_T$ decays exponentially. Therefore, when
$\eta_T$ becomes as small as $1/3$ the crystal loses its stability with respect to the hexatic phase The brackets $<>_{replica}$ mean averaging over ten replicas of the same
system, but with different positions of the pinning centers.

The orientational correlation function is defined as

\begin{equation}
g_6(r)=\left< \frac{\left<\psi_6({\bf r})\psi_6^*({\bf 0})\right>}{g(r)} \right>_{replica},
\label{g6}
\end{equation}
where $ \psi_6({\bf r_i})=\frac{1}{n(i)}\sum_{j=1}^{n(i)}
e^{in\theta_{ij}}$ is a bond orientational order parameter. Here $n(i)$ is the number of the nearest neighbors of
the $i-$th particle and $\theta_{ij}$ is the angle between the
bond between  $i-$th particle and its $j-$th neighbor and an
arbitrary axis.  In the hexatic phase the long-range behavior of $g_6(r)$ has the form
$g_6(r)\propto r^{-\eta_6}$ with $\eta_6 \leq \frac{1}{4}$
\cite{halpnel1, halpnel2}. Therefore, when $\eta_6$ becomes as
small as $1/4$ the hexatic phase loses its stability with respect
to isotropic liquid.

All simulations were performed using the LAMMPS simulation package
\cite{lammps}.

\section{Results and Discussion}

\subsection{The systems without pinning and with Gaussian pinning and well depth $1.0$}

\begin{figure}
\includegraphics[width=8cm]{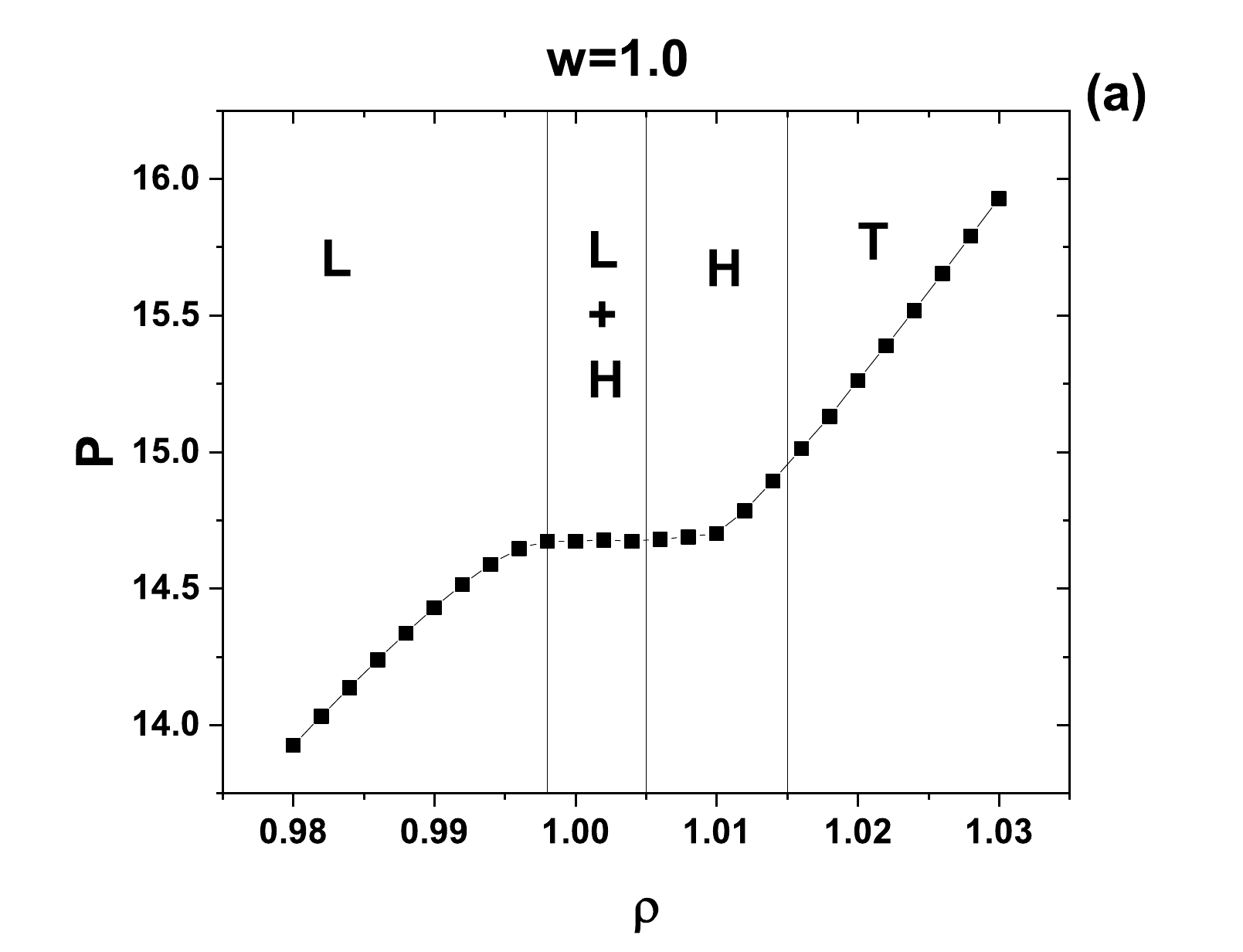}%

\includegraphics[width=8cm]{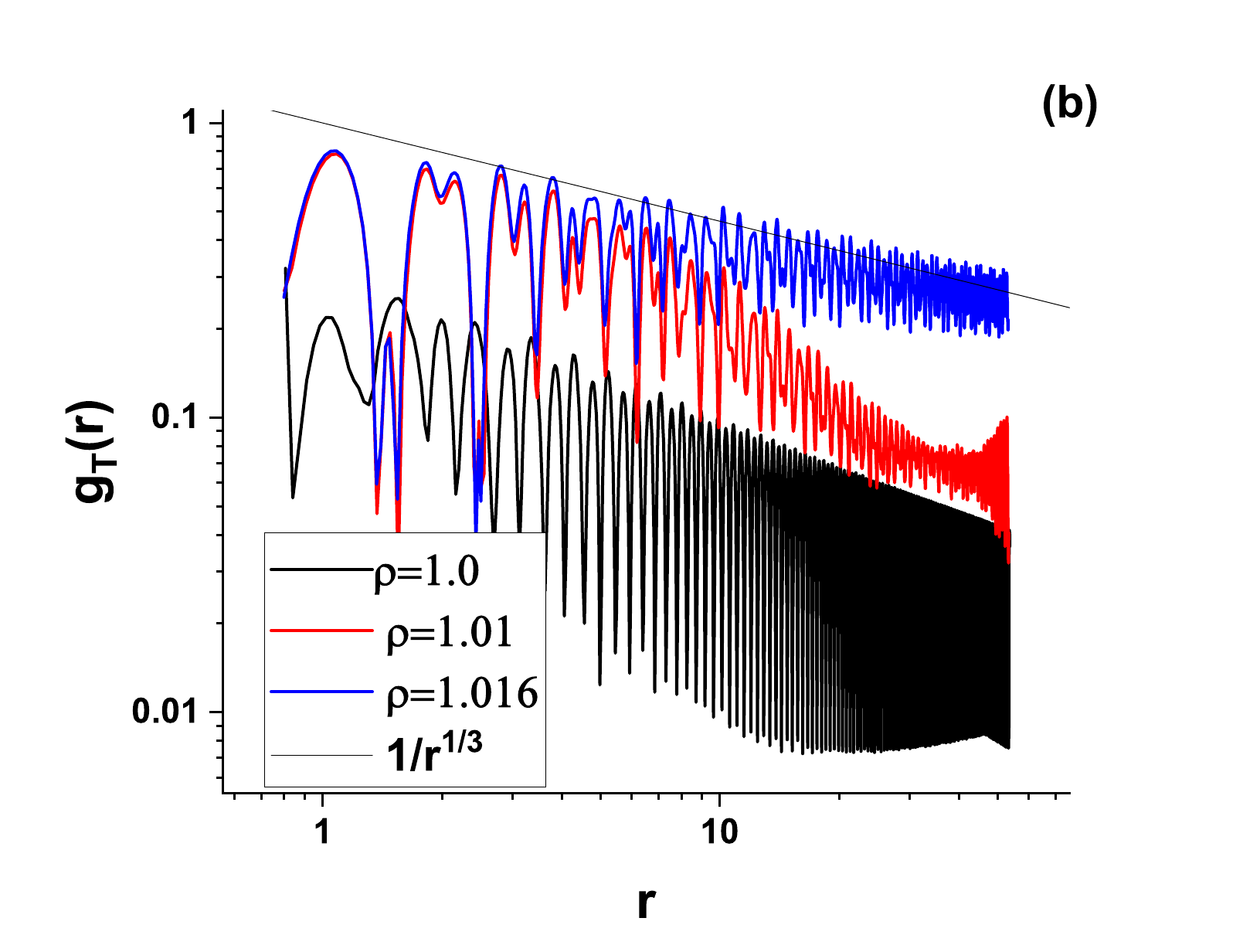}%

\includegraphics[width=8cm]{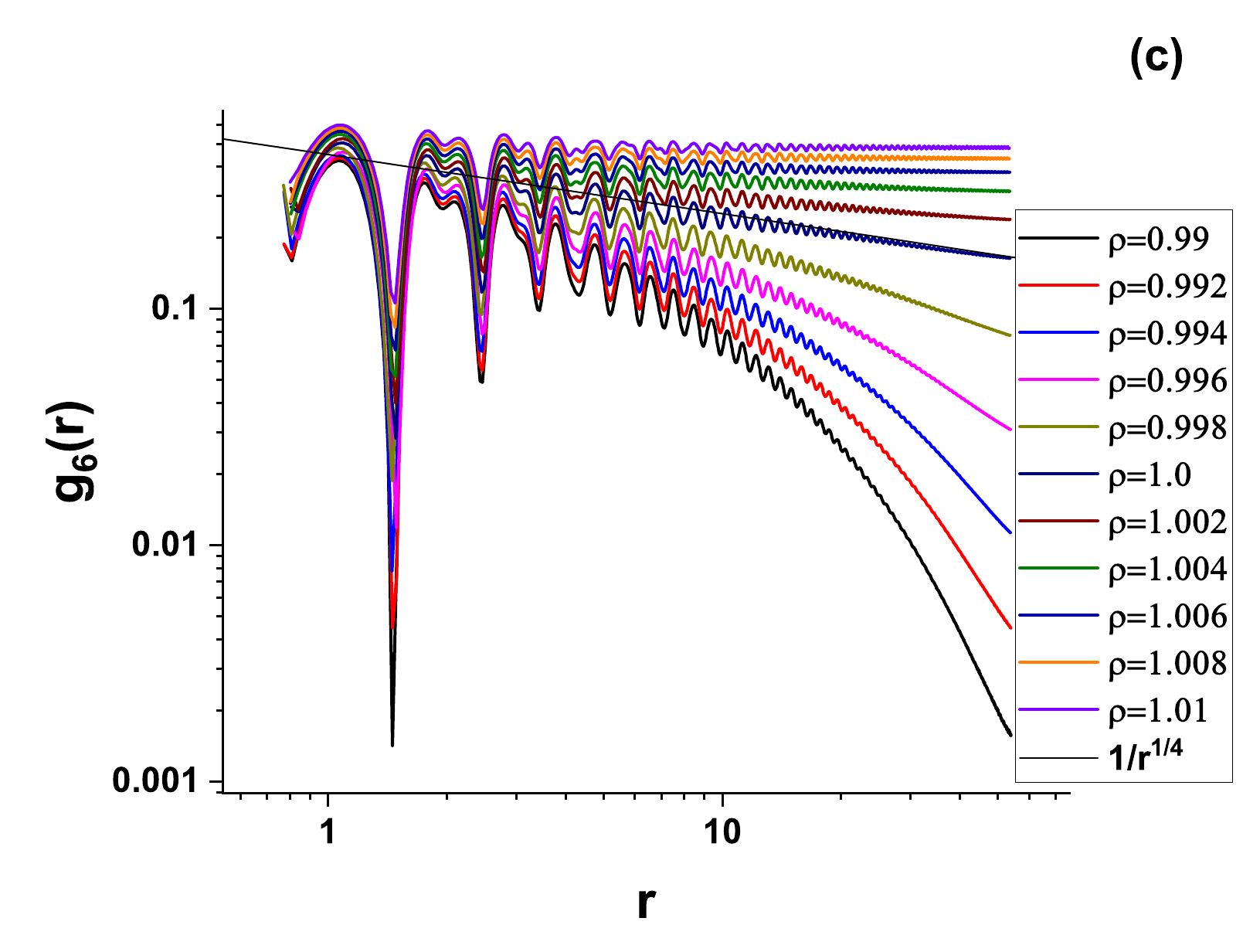}%

\caption{\label{w1} The system with Gaussian pinning and well depth $w=1.0$. (a)
The equation of state in the melting region. The letters denote the equilibrium phase in the system. $L$ is isotropic
liquid, $L+H$ is a liquid-hexatic two-phase region, $H$ is a
hexatic phase and $T$ is a triangular crystal. (b) The
translational correlation functions of the same system. (c) The
orientational correlation functions of the same system.}
\end{figure}

We do not find any difference in the behavior of
the systems without pinning and with Gaussian pinning and well depth $w=1.0$ due to
the comparability of the well depth with the simulation
temperature. For this reason we demonstrate only the results for
the system with Gaussian pinning and $w=1.0$.

The equation of state of the system with Gaussian pinning and $w=1.0$ is shown in Fig.
\ref{w1} (a). As is seen from this figure, the equation of state
demonstrates the Mayer-Wood loop, i.e. a first-order transition takes place
in the system. Panels (b) and (c) of the same figure show the
translational and orientational correlation functions. Based on
the behavior of these functions behavior we conclude that the system melts
according to the BK scenario, a continuous BKT transition from
crystal to hexatic and a first-order one from hexatic to isotropic
liquid. In this case the liquid-hexatic transition boundary is
obtained using the Maxwell construction. The transition points are exactly the same as for the
system without pinning and with Gaussian pinning and $w=1.0$: $\rho_{sh}=1.015$ for the
continuous transition from solid to hexatic, $\rho_{hl}=1.005$ and
$\rho_l=0.998$ for the coexistence densities of hexatic and
isotropic liquid respectively. Exactly the same transition points
are found in the system without pinning \cite{n12}.


\subsection{The system with Gaussian pinning and well depth $10.0$}

\begin{figure}
\includegraphics[width=8cm]{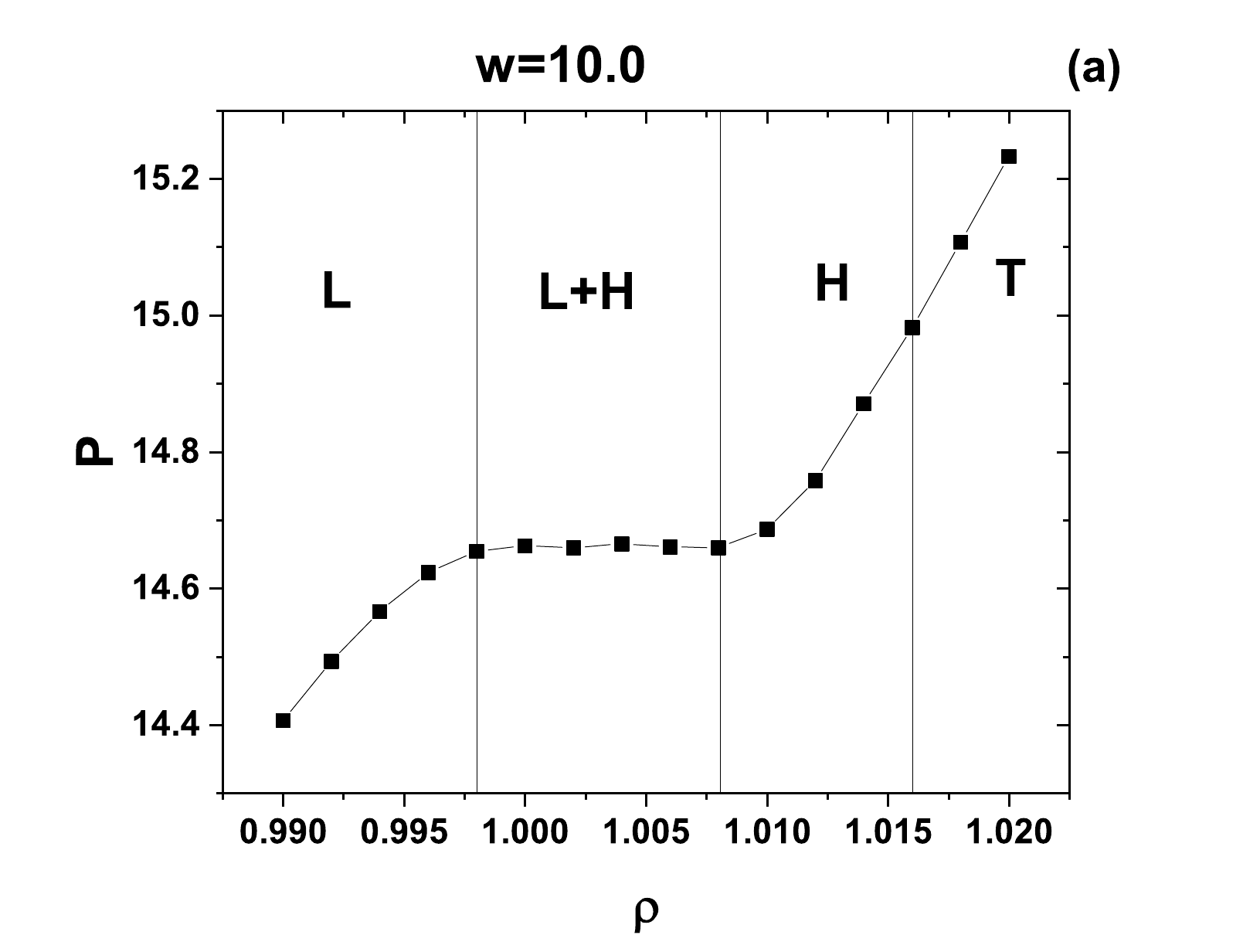}%

\includegraphics[width=8cm]{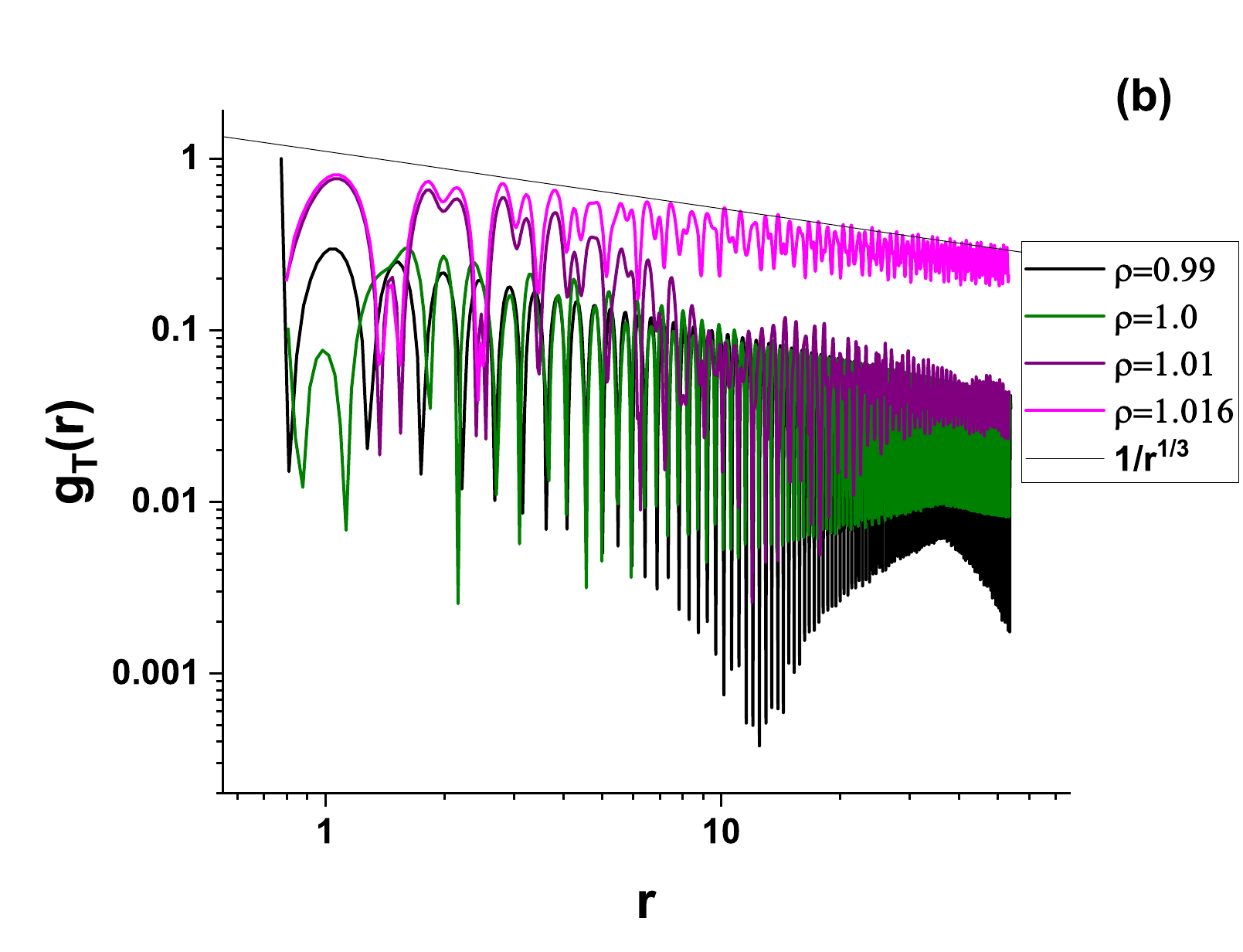}%

\includegraphics[width=8cm]{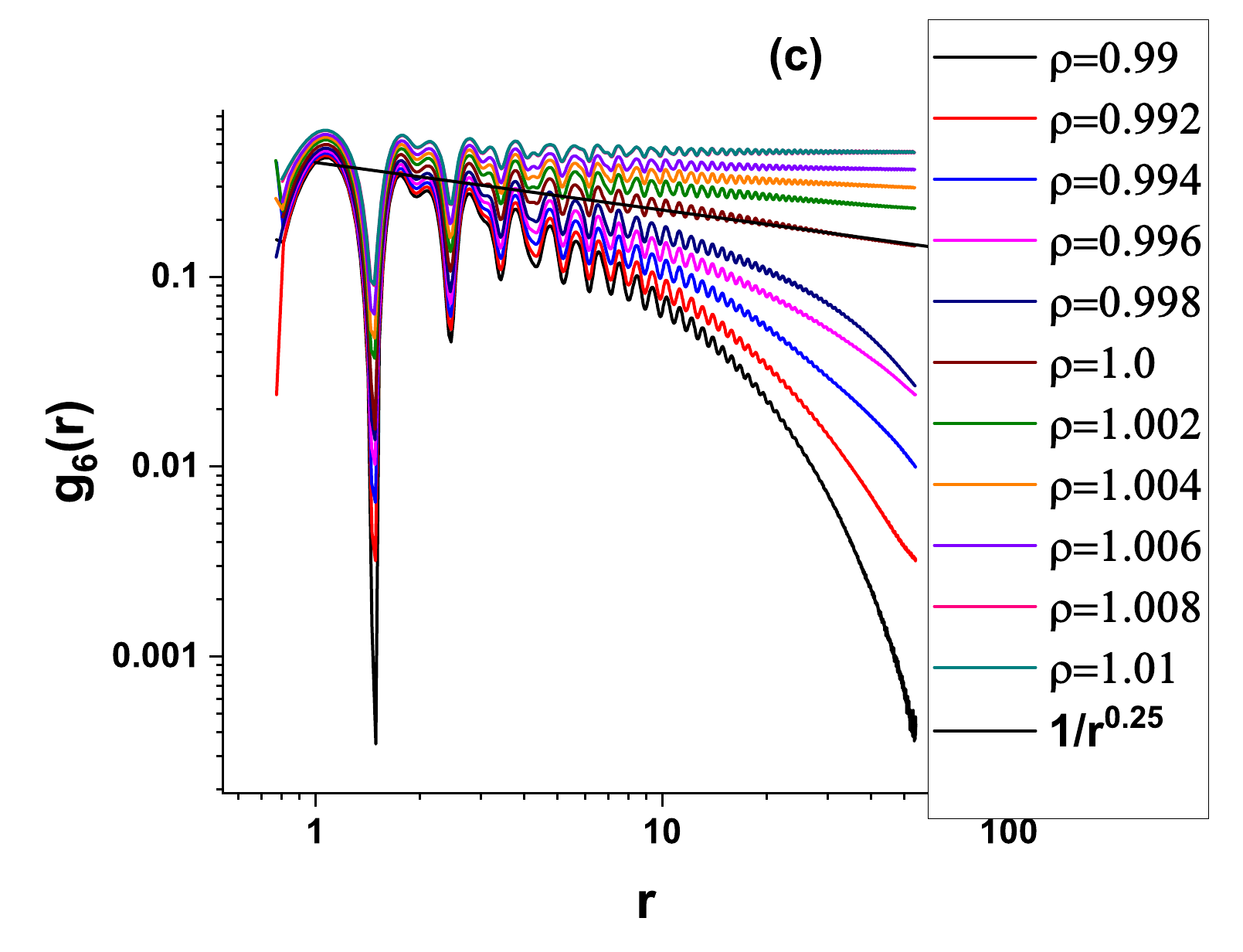}%
\caption{\label{w10} The system with Gaussian pinning and well depth $w=10.0$. (a)
The equation of state in the melting region.
The letters denote the equilibrium phase in the system. $L$ is isotropic
liquid, $L+H$ is a liquid-hexatic two-phase region, $H$ is a
hexatic phase and $T$ is a triangular crystal. (b) The
translational correlation functions of the same system. (c) The
orientational correlation functions of the same system.}
\end{figure}

\begin{figure}
\includegraphics[width=8cm]{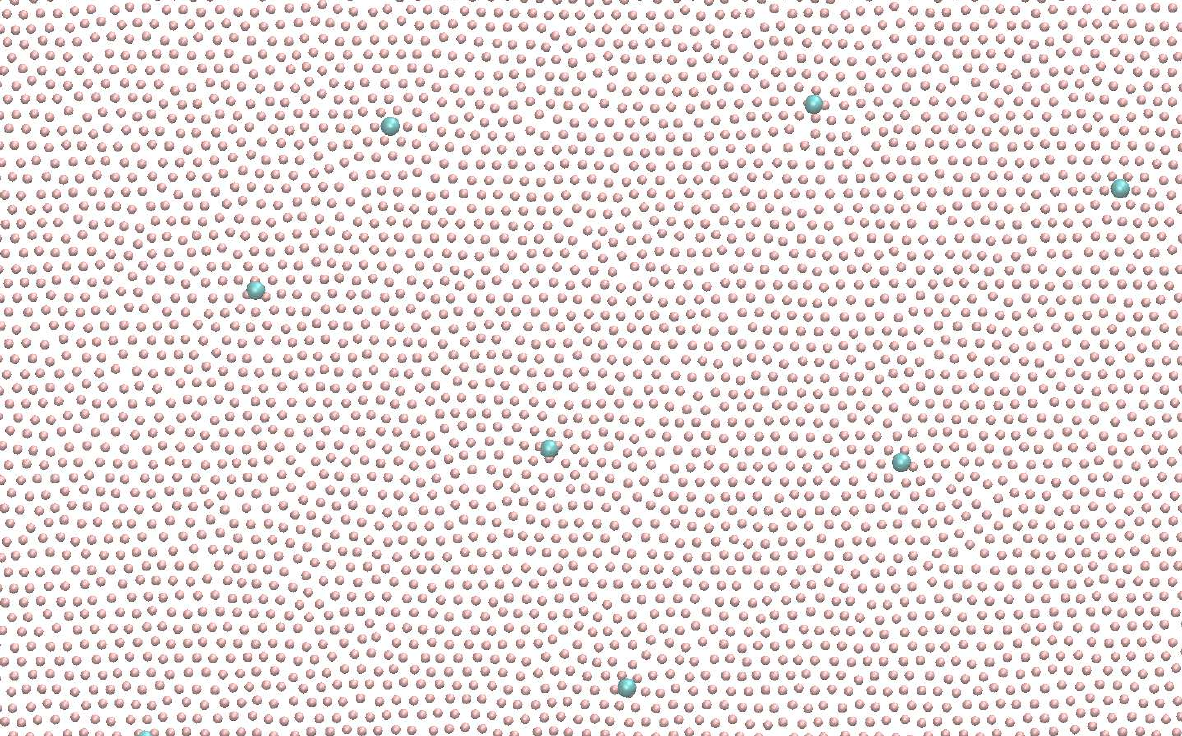}%
\caption{\label{w10s} A snapshot of the system with $\rho=1.016$ and Gaussian pinning well depth $w=10.0$. The small pink circles are particles and the big cyan circles are pinning centers.}
\end{figure}

The melting scenario preserves if the well depth becomes $w=10.0$
(Fig. \ref{w10}(a)-(c)). However, in this case the density range
where the hexatic phase is stable becomes smaller, while the
two-phase region enlarges. The densities of phase transitions
are $\rho_{sh}=1.016$ for the solid to hexatic transition (BKT
type melting) and $\rho_{hl}=1.0086$ and $\rho_{l}=0.998$ for the
coexistence densities of hexatic and isotropic liquid
respectively.

Figure \ref{w10s} shows a snapshot of the system at
the density $\rho=1.016$ which belongs to the region of stability
of the crystalline phase  but near the solid-hexatic transition
point. It is seen from this figure that dense stable clusters
appear in the solid phase. They mainly consist of only four
particles due to the small value of $r_{cg}=1.27$. Such defects lower the effective density of the system and lead to a density increase in the BKT solid-hexatic transition.


\subsection{The system with Gaussian pinning and well depth $50.0$}

\begin{figure}
\includegraphics[width=8cm]{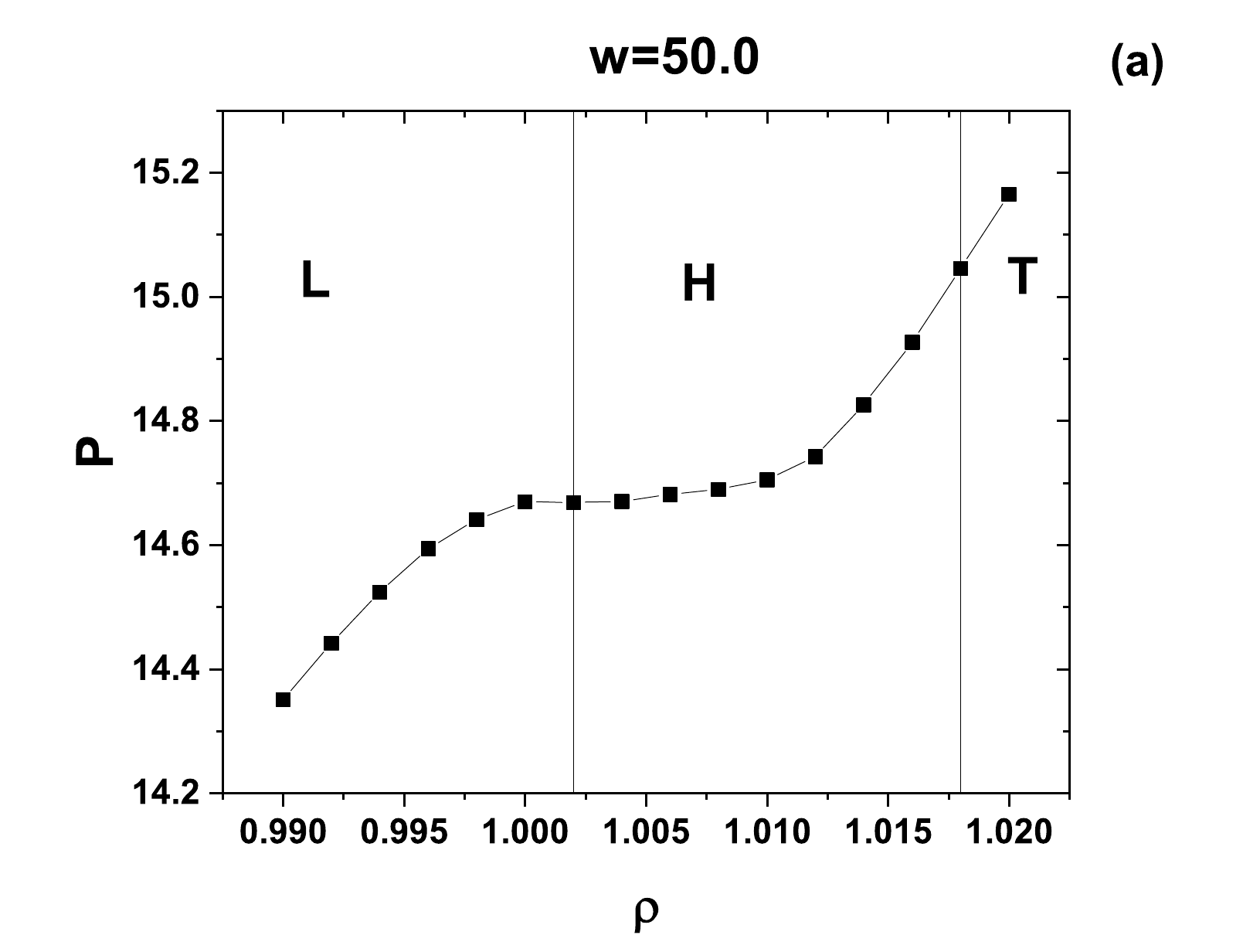}%

\includegraphics[width=8cm]{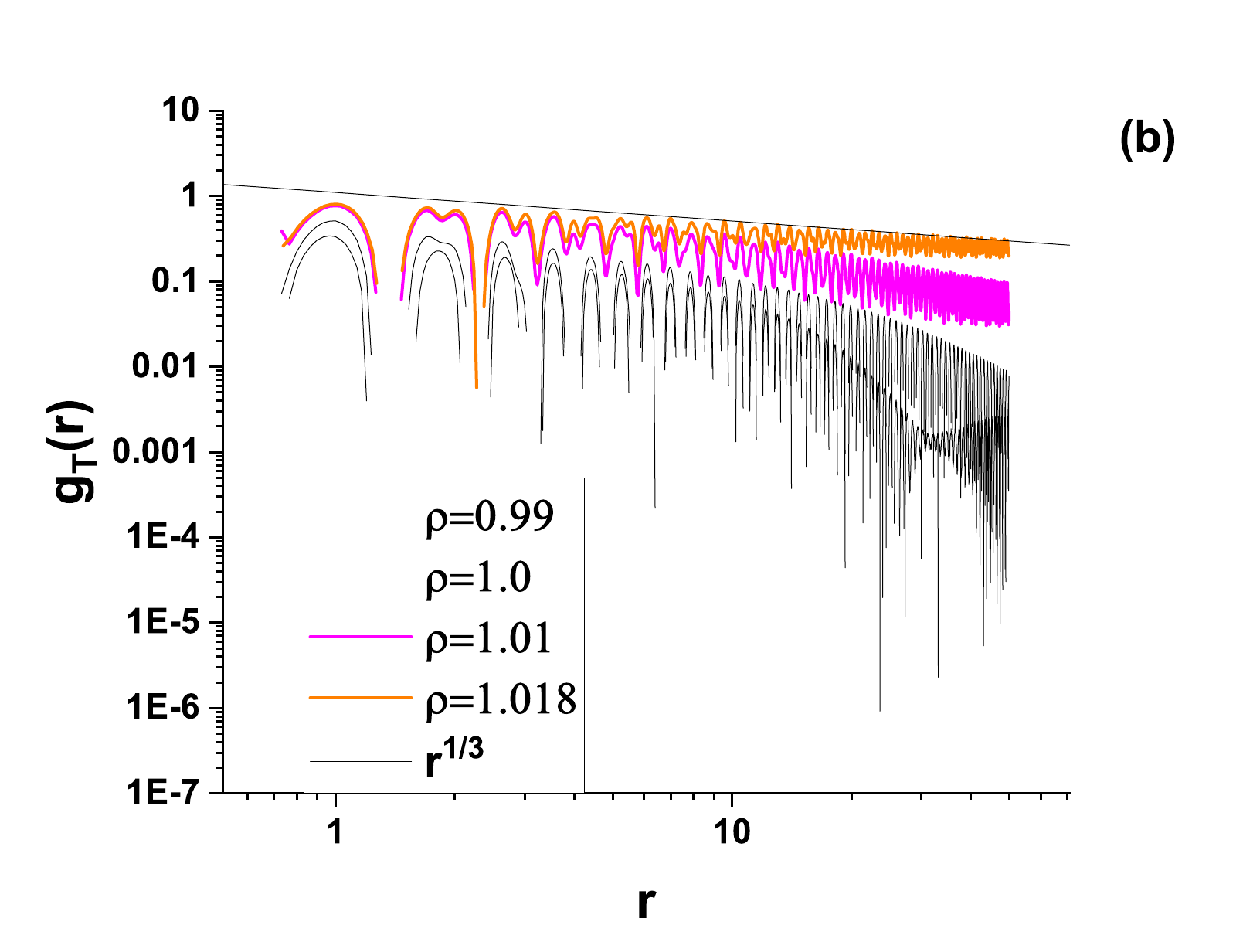}%

\includegraphics[width=8cm]{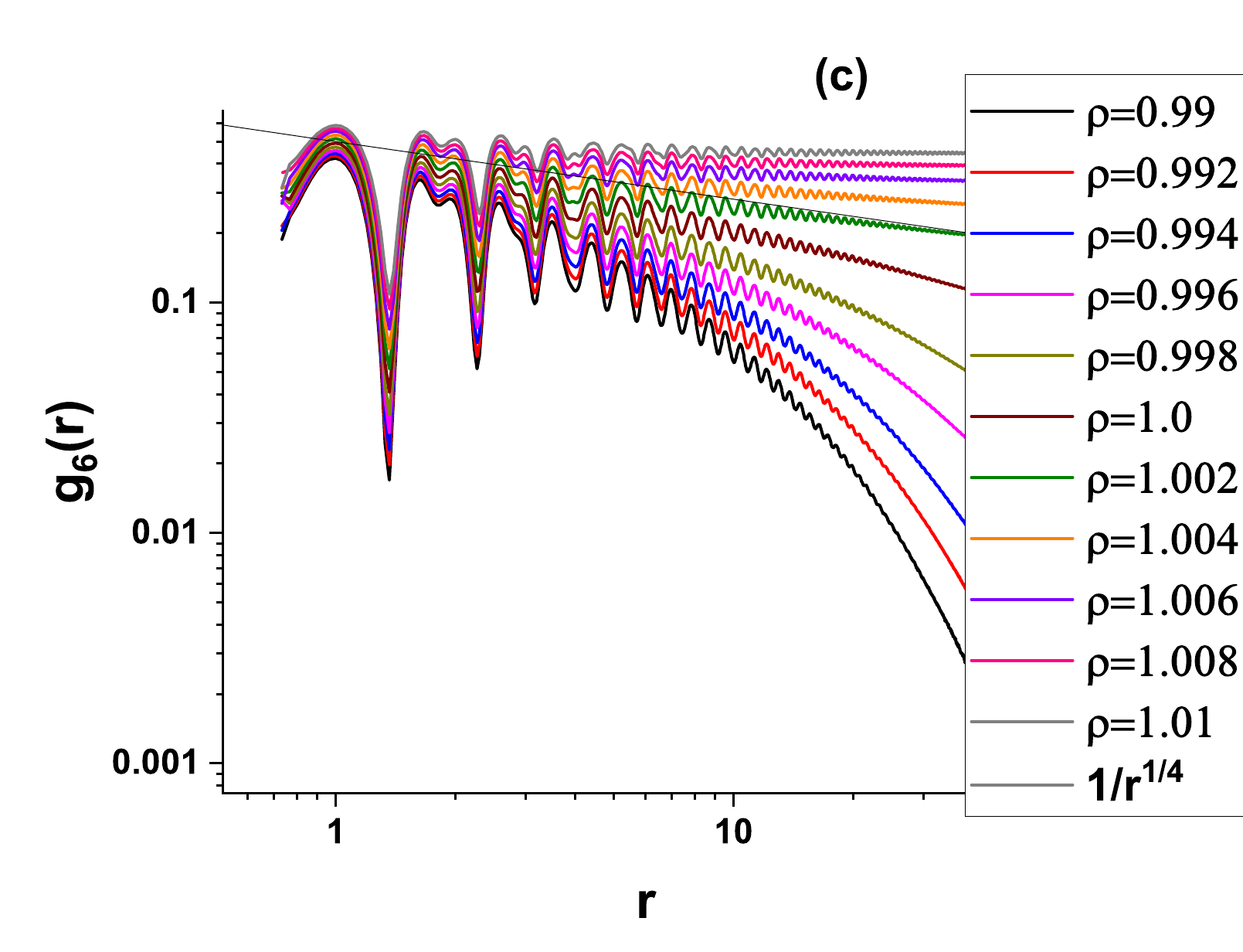}%

\caption{\label{w50} The system with Gaussian pinning and well depth $w=50.0$. (a)
The equation of state in the melting region.
The letters denote the equilibrium phase in the system. $L$ is isotropic
liquid, $H$ is a hexatic phase and $T$ is a triangular crystal.
(b) The translational correlation functions of the same system.
(c) The orientational correlation functions of the same system.}
\end{figure}

\begin{figure}

\includegraphics[width=8cm]{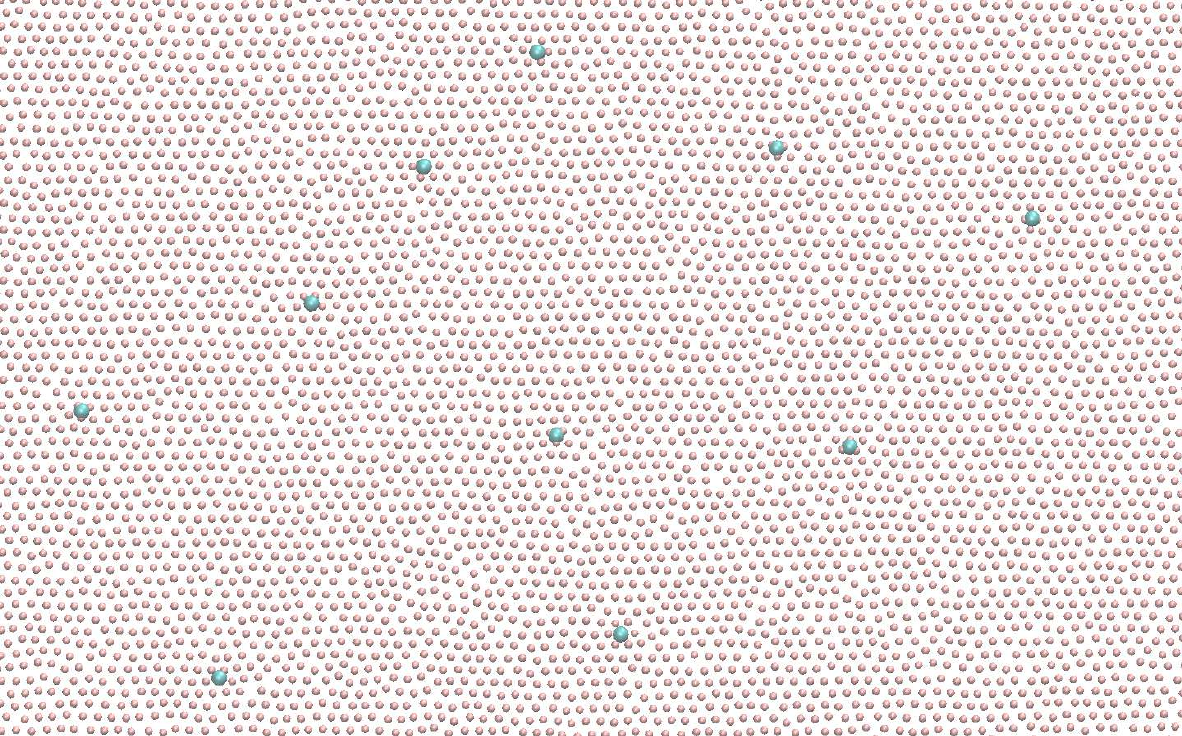}%

\caption{\label{w50s} A snapshot of the system with $\rho=1.02$ and Gaussian pinning well depth $w=50.0$. The small pink circles are particles and the big cyan circles are pinning centers.}
\end{figure}

In the case of $w=50.0$ the melting scenario changes. From Fig.
\ref{w50} (a) one can see that the equation of state does not
demonstrate the Mayer-Wood loop, i.e. no first-order phase
transition takes place in this system. The melting scenario
occurs according to the BKTHNY theory. The transition points are
determined from the translational and orientational correlation
functions shown in Fig. \ref{w50} (b) and (c) respectively. The
density of melting of the crystal into the hexatic phase is
$\rho_{sh}=1.018$ and the transition from the hexatic phase to
isotropic liquid takes place at $\rho=1.002$.

Figure \ref{w50s} shows a snapshot of the system
at $\rho=1.02$ which belongs to the region of stability of the
crystal. In this case, one can see some defects in the crystal
structure of the system: there is some concentration in the
vicinity of the pinning centers and rarefication far from them.
The clusters consist of four, five, six and seven particles. Most
probably, these defects are responsible for the change of the
melting scenario of the system. As it was mentioned above, such defects
lower the effective density of the system and lead to an increase in the density of the BKT solid-hexatic transition.

\subsection{The system with Gaussian pinning and well depth $100.0$}

\begin{figure}

\includegraphics[width=8cm]{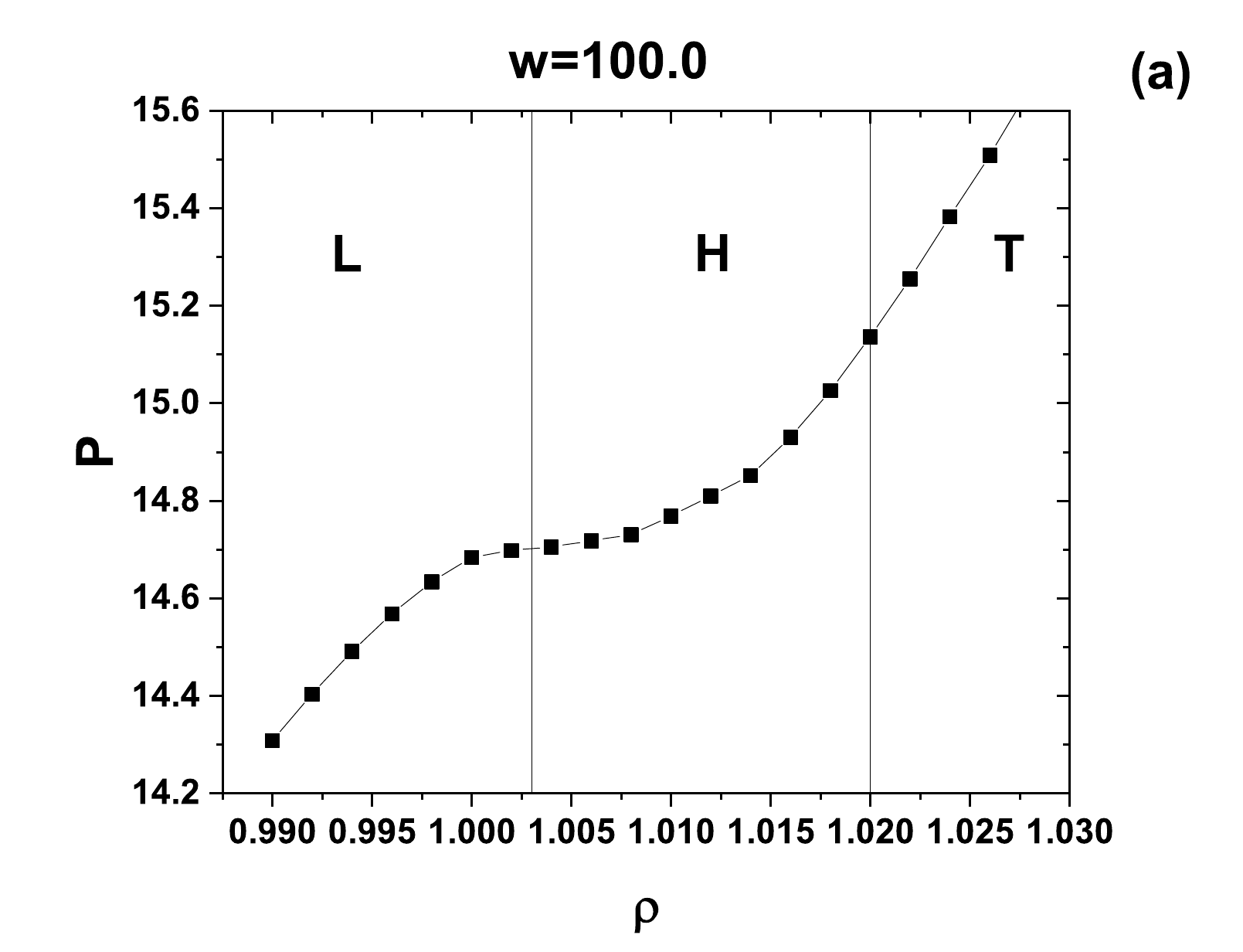}%

\includegraphics[width=8cm]{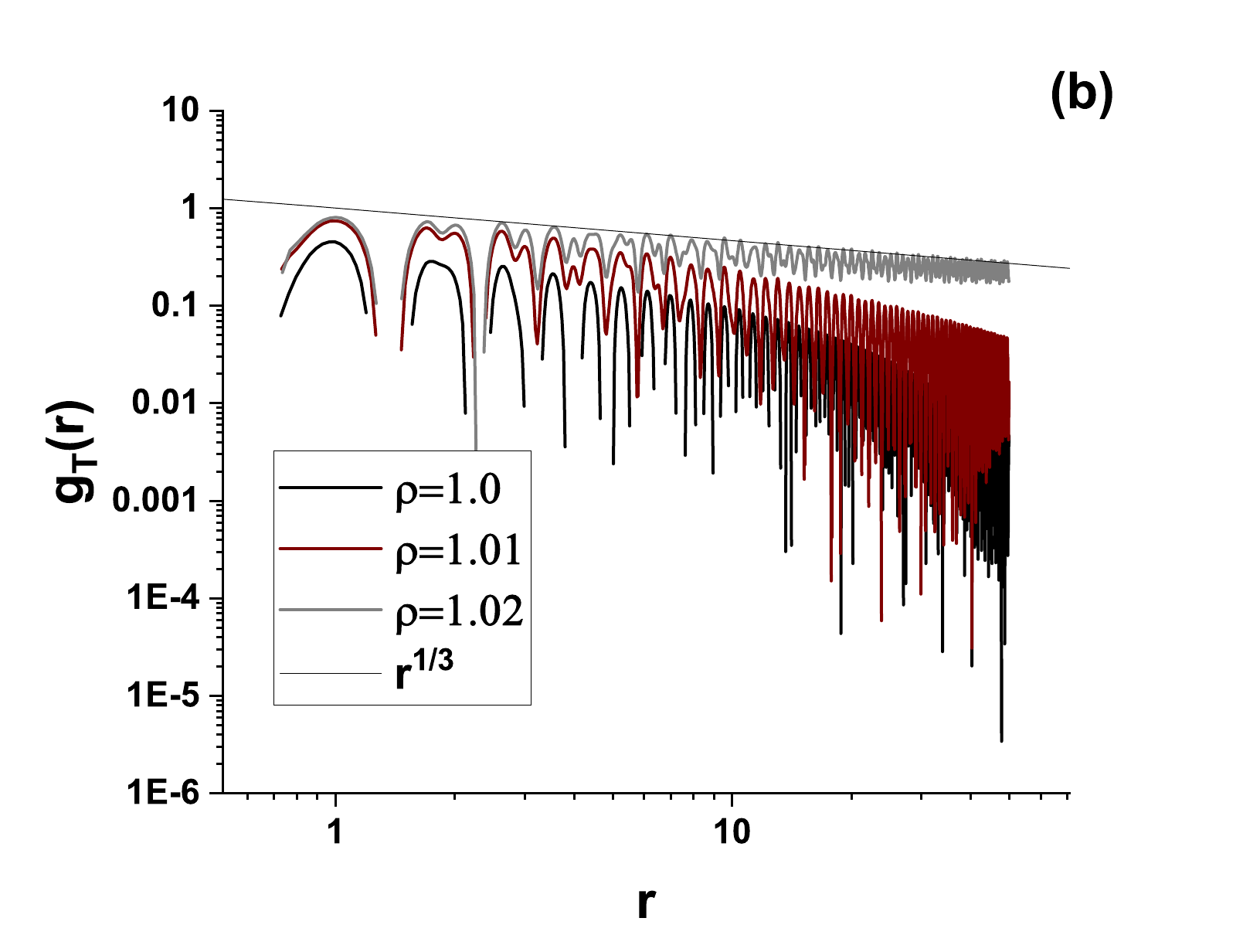}%

\includegraphics[width=8cm]{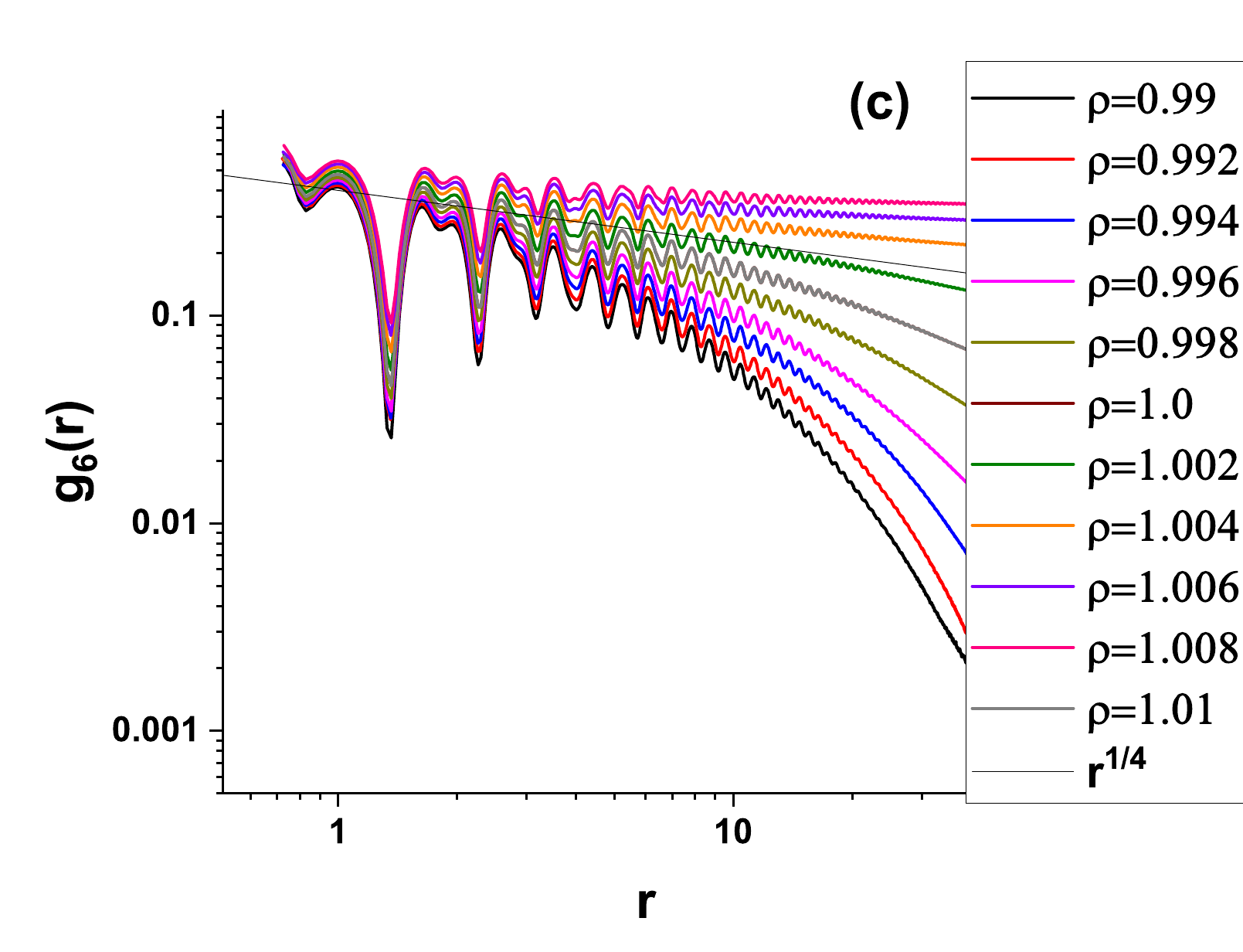}%

\caption{\label{w100} The system with Gaussian pinning and well depth $w=100.0$.
(a) The equation of state in the melting region.
The letters denote the equilibrium phase in the system. $L$ is isotropic
liquid, $H$ is a hexatic phase and $T$ is a triangular crystal.
(b) The translational correlation functions of the same system.
(c) The orientational correlation functions of the same system.}
\end{figure}

\begin{figure}

\includegraphics[width=8cm]{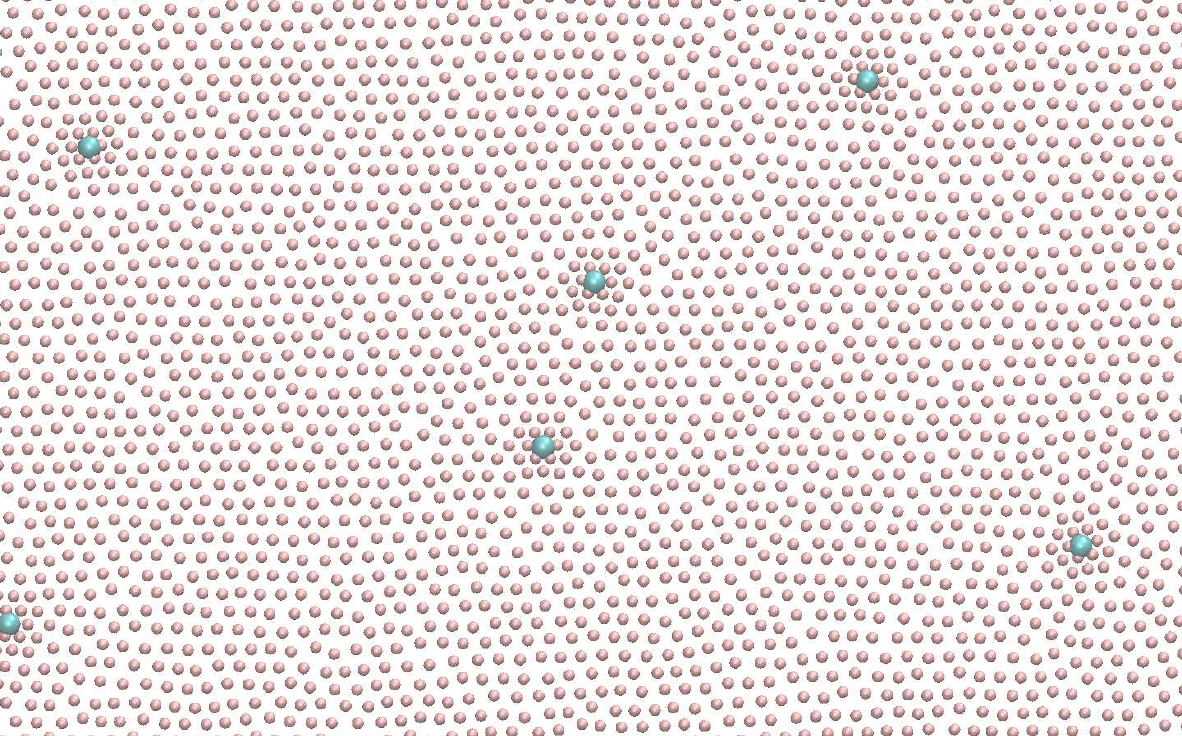}%

\caption{\label{w100s} A snapshot of the system with $\rho=1.03$ and Gaussian pinning well depth $w=100.0$. The small pink circles are particles and the big cyan circles are pinning centers.}
\end{figure}

Next we consider the system with Gaussian pinning well depth $w=100.0$. Figure \ref{w100} (a) shows the equation of state of this system. No Mayer-Wood loop is observed, therefore the melting scenario remains according to the BKTHNY theory. The transition points are determined from the correlation functions: $\rho_{sh}=1.02$ and $\rho_{lh}=1.003$ for the transition from the crystal to the hexatic phase and from the hexatic phase to liquid respectively.

At the same time at this large well depth we
see strong influence of $r_{cg}$ on cluster formation. A
snapshot of the system at density $\rho=1.03$ (the crystalline
phase) is shown in Fig. \ref{w100s}. One can see that the pinning
centers become the centers of crystallization. These clusters are
significantly larger than in the case of $w=10.0$ and $50.0$: they
involve about two coordination spheres. At larger distances, no
apparent effects of the pinning centers are observed, despite the decrease in the mean density between these centers.

\subsection{The system with Gaussian pinning and well depth $200.0$}

\begin{figure}

\includegraphics[width=8cm]{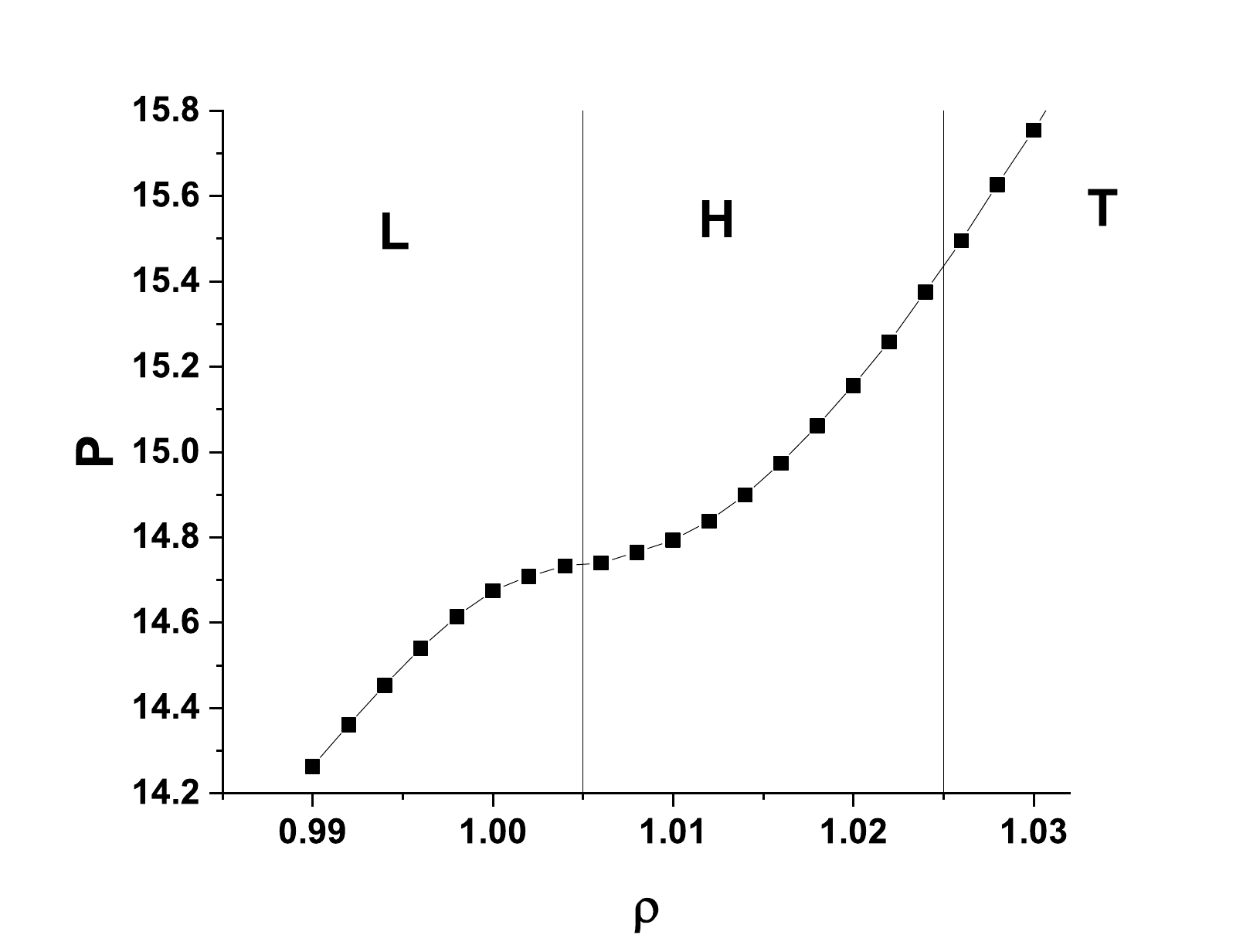}%

\includegraphics[width=8cm]{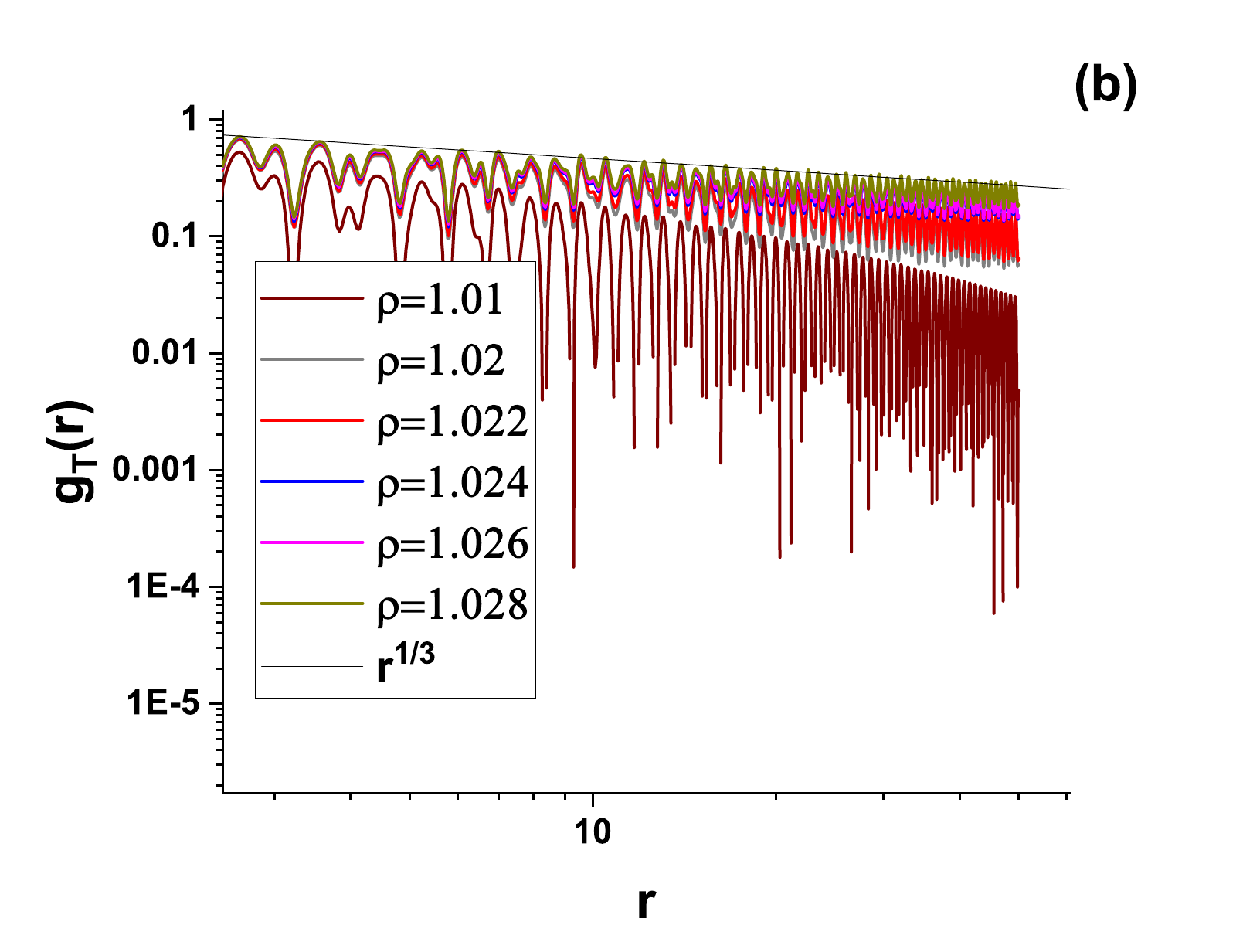}%

\includegraphics[width=8cm]{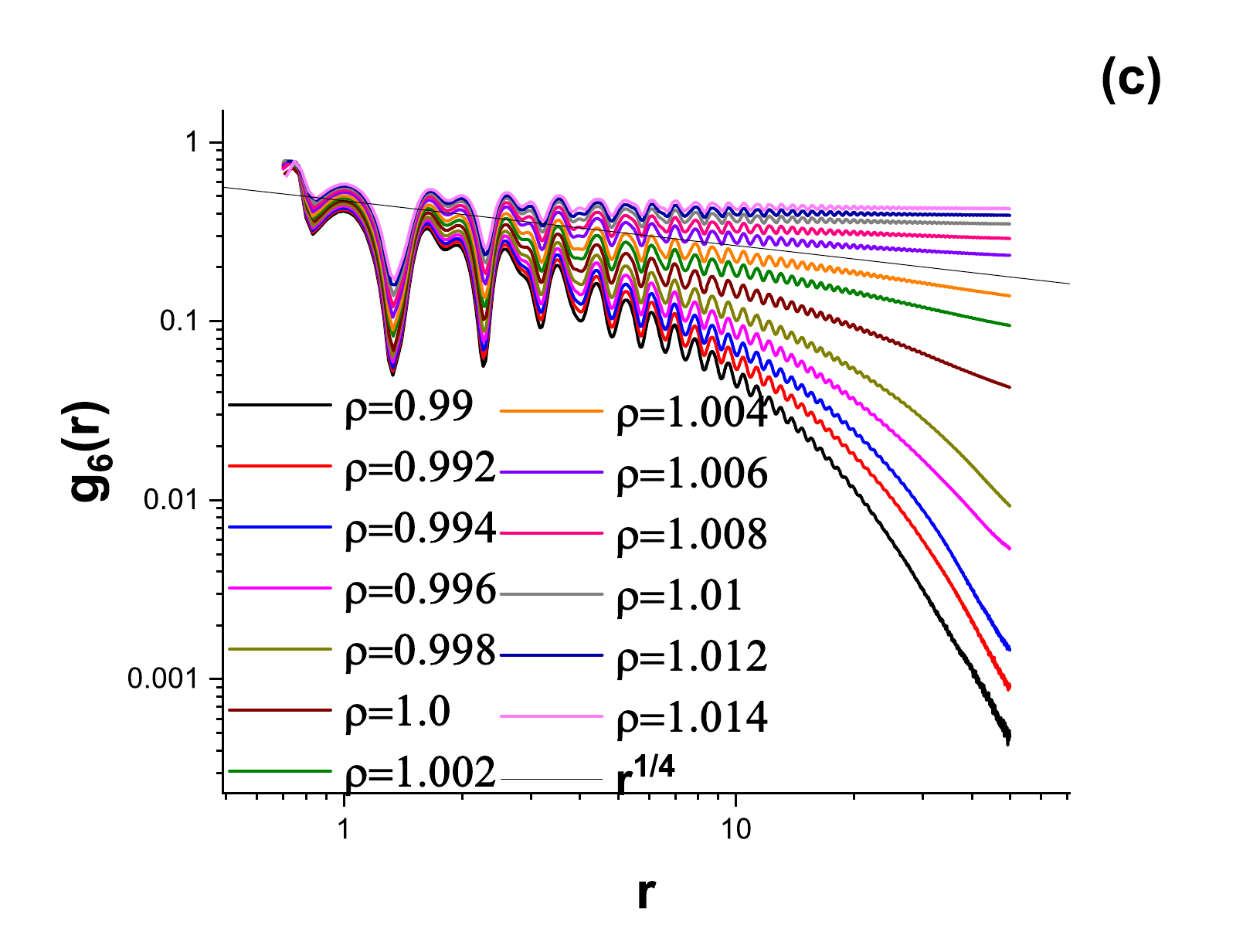}%

\caption{\label{w200} The system with Gaussian pinning and well depth $w=200.0$.
(a) The equation of state in the melting region.
The letters denote the equilibrium phase in the system. $L$ is isotropic
liquid, $H$ is a hexatic phase and $T$ is a triangular crystal.
(b) The translational correlation functions of the same system.
(c) The orientational correlation functions of the same system.}
\end{figure}

\begin{figure}

\includegraphics[width=8cm]{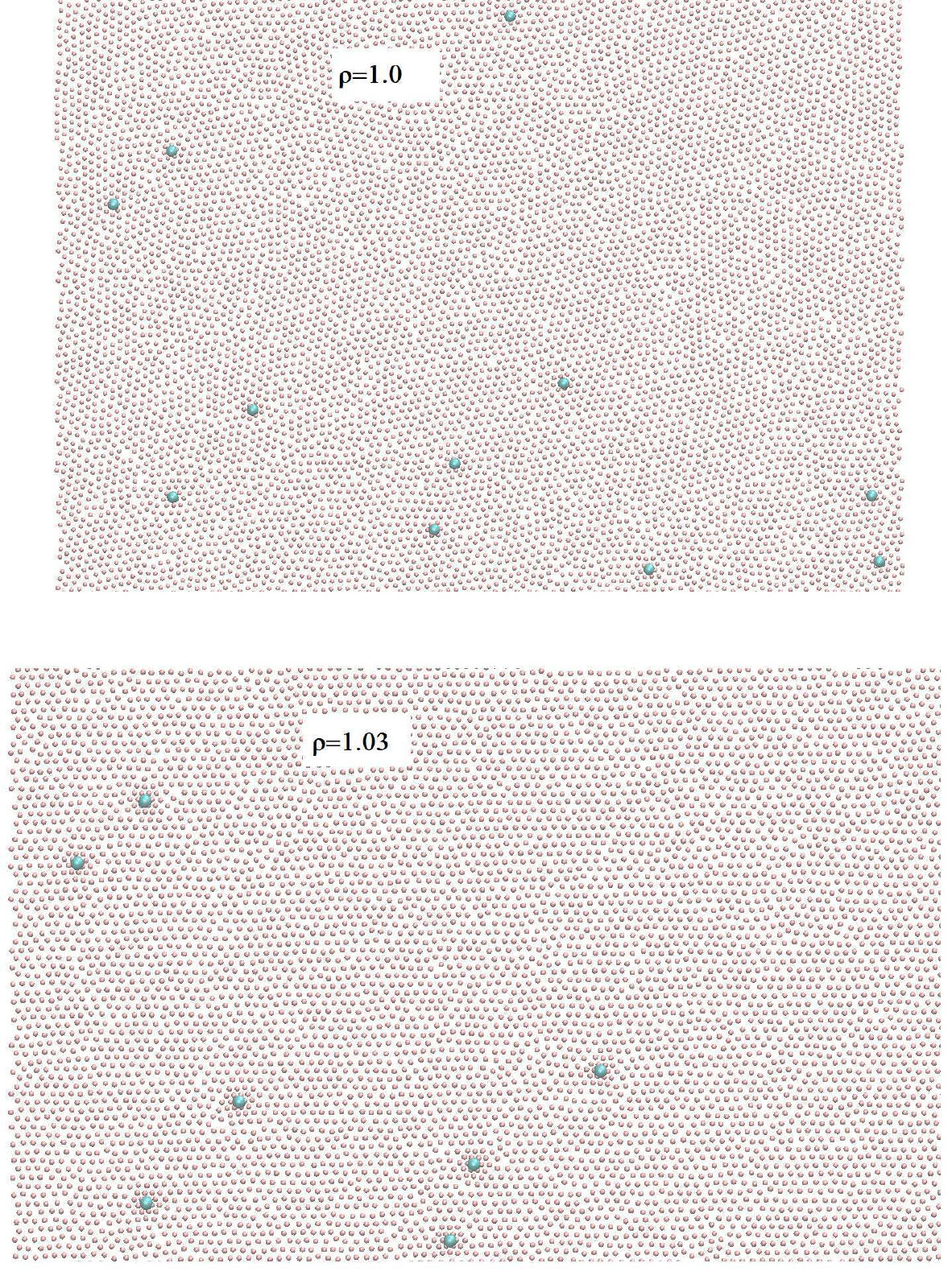}%

\caption{\label{w200s} Snapshots of the systems with $\rho=1.0$
and $\rho=1.03$, and Gaussian pinning well depth $w=200.0$. The small pink circles are
particles and the big cyan circles are pinning centers.}
\end{figure}

The deepest well considered in this study is $w=200.0$. The
results are shown in Fig. \ref{w200}. The equation of state of
this system does not demonstrate the Mayer-Wood loop, therefore the
system melts according to the BKTHNY scenario. The boundaries of
phases are determined via analysis of the correlation functions. The
resulting densities are $\rho_{lh}=1.005$ and $\rho_{sh}=1.025$
for the hexatic to liquid and solid to hexatic transitions
respectively.

In Fig. \ref{w200s} we show snapshots of the systems with Gaussian pinning well depth $w=200.0$ at densities $\rho=1.0$ (the liquid phase) and $\rho=1.03$ (the triangular crystal). Once again, we see that the particles in the system concentrate in the vicinity of the pinning centers. But in both liquid and crystalline phases, the cluster sizes around the pinning centers are two coordination spheres only. Therefore, even if the well depth is rather large, its influence
on the overall structure of the system is relatively modest.
This once again underlines the conclusion that
$r_{cg}$, and not the depth of the well, is the determining factor
in the formation of clusters in the vicinity of pinning centers.
This result is similar to the result for $w=100.0$, since their
values of $r_{cg}$ differ slightly.

\subsection{Comparison of different well depths}

\begin{figure}

\includegraphics[width=8cm]{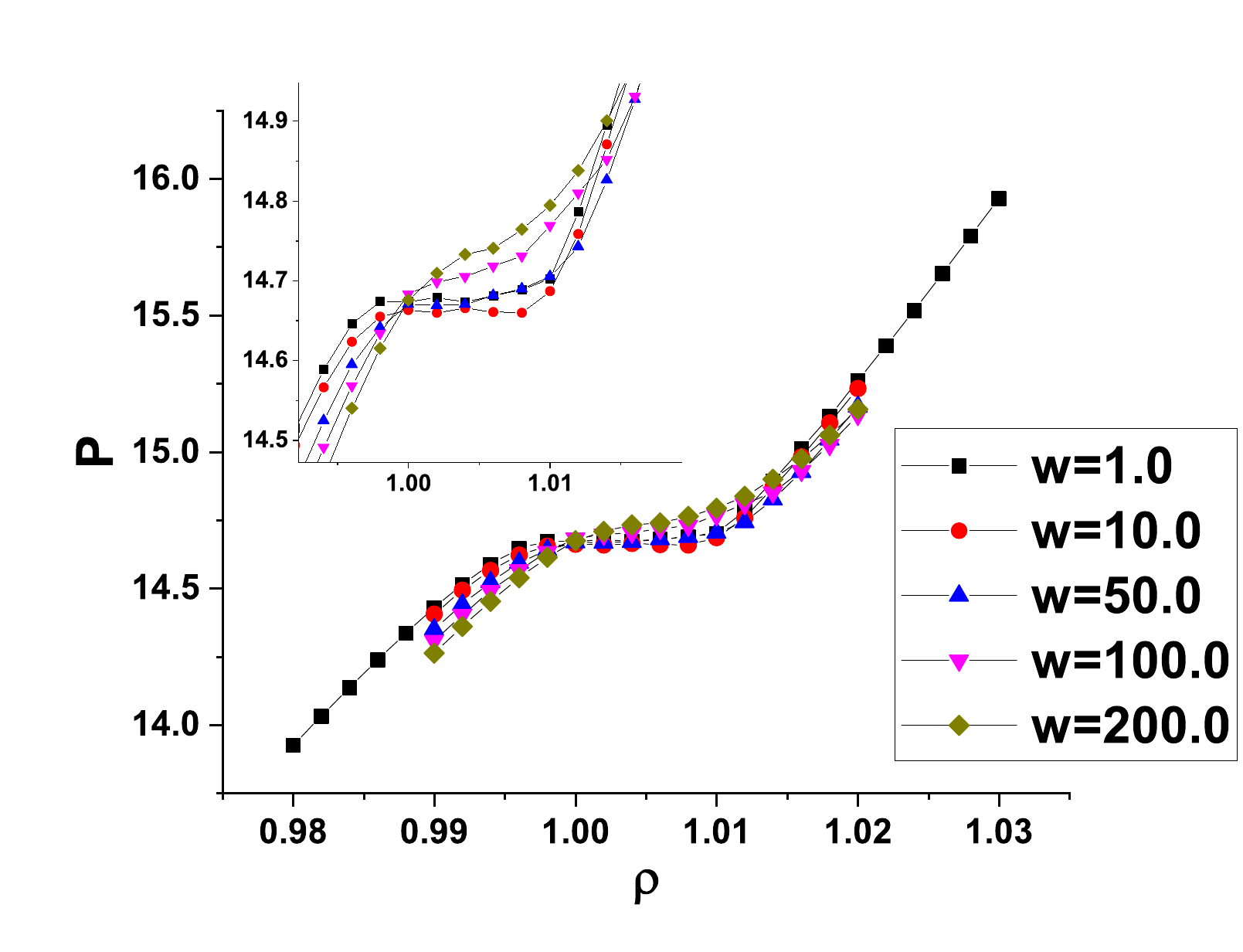}%

\caption{\label{all} The equations of state of all simulated
systems. The inset enlarges the region of phase transition.}
\end{figure}

Finally, we compare the equations of state of all considered
systems (Fig. \ref{all}). An increase in the
overall melting density with an increase in $w$ is associated with a
decrease in the average effective density of the system due to the
movement of part of the particles to the pinning
centers and due to an increase in the effective size of these
clusters.

\begin{figure}

\includegraphics[width=8cm]{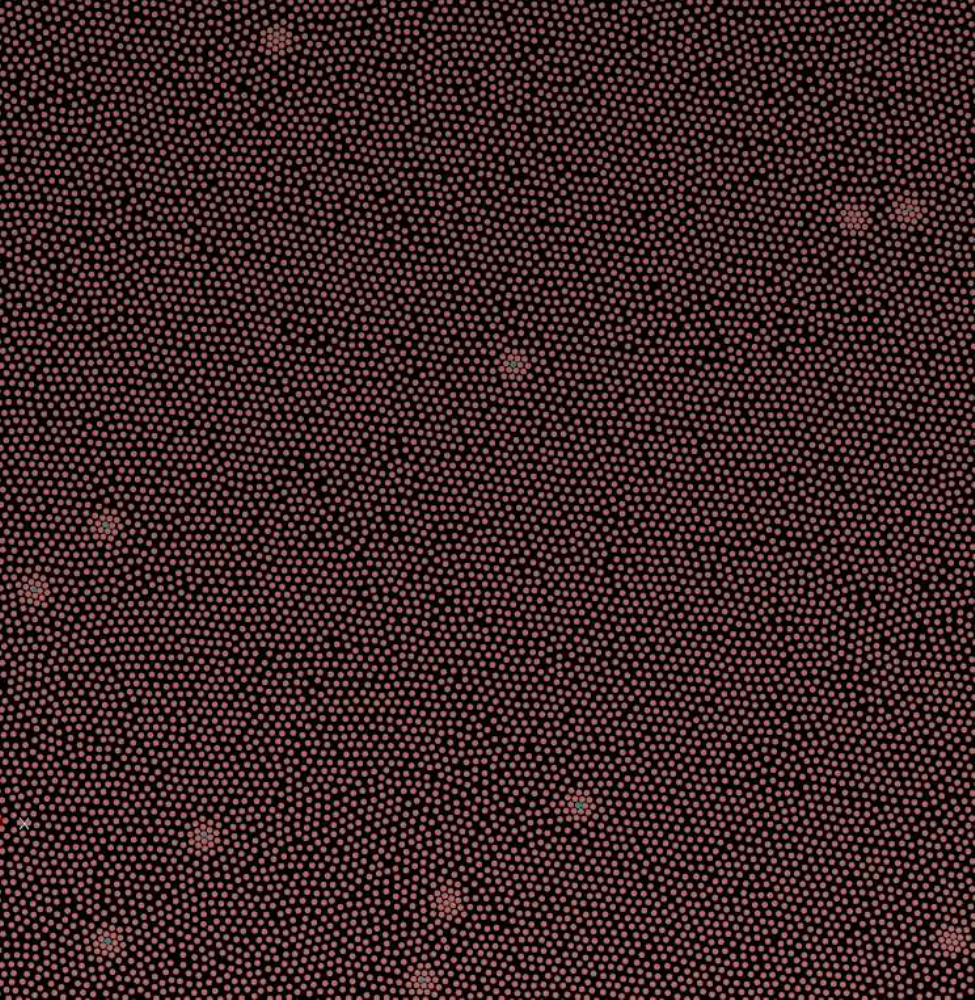}%

\caption{\label{w200h} A snapshot of the hexatic phase at
$\rho=1.01$ and Gaussian pinning well depth $w=200.0$. The small pink circles are particles
and the big cyan circles are pinning centers.}
\end{figure}

\begin{figure}

\includegraphics[width=10cm]{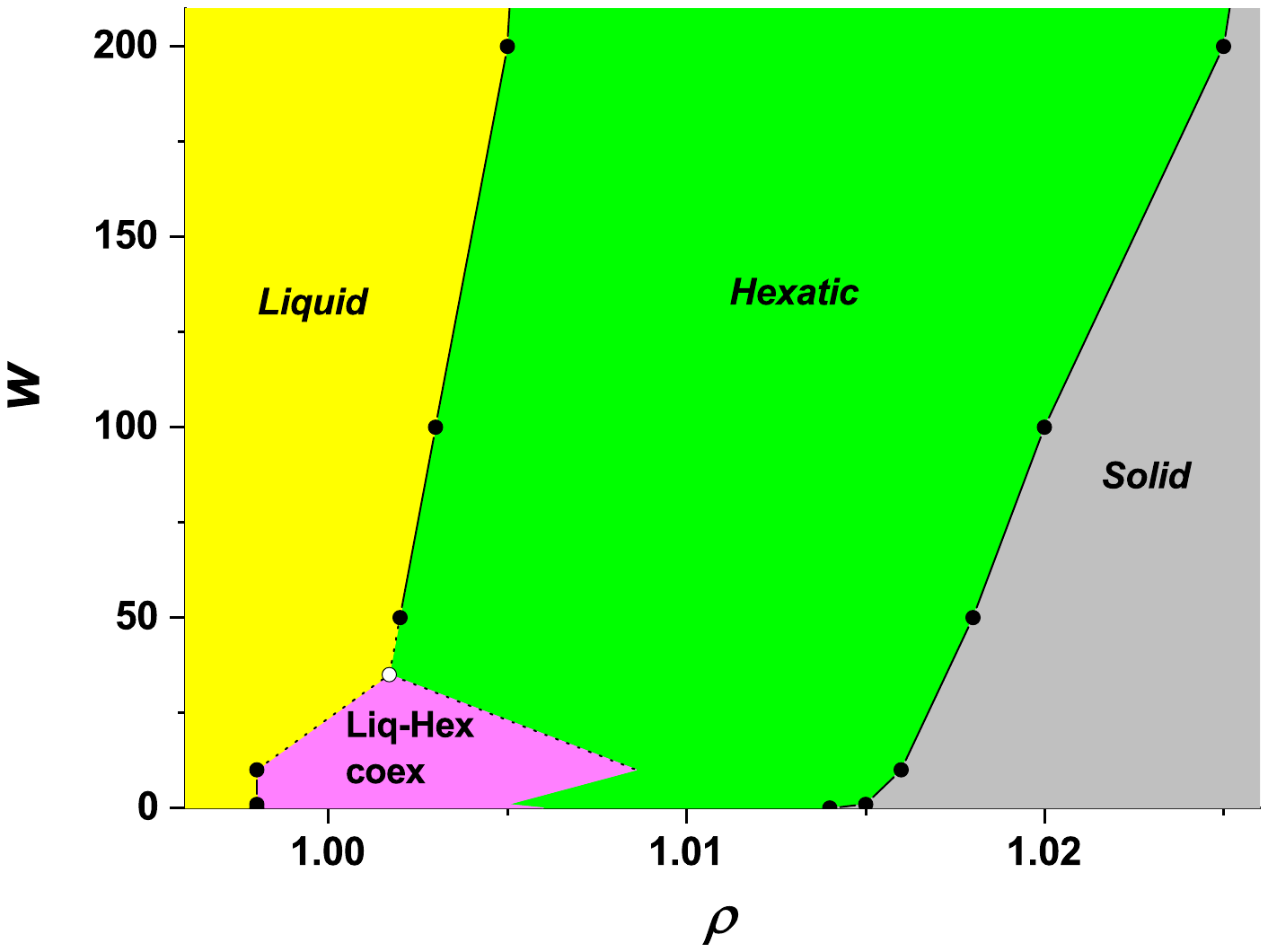}%

\caption{\label{phd} The phase diagram of the dependence of the
well depth on density. }
\end{figure}

As can be seen from Fig. \ref{all}, the equation of state of the
system in the presence of pinning for $w=10.0$ uniformly went down
in pressure compared to the equation of state of the system for
$w=1.0$ or a system without pinning. The decrease in pressure is
associated with a decrease in the average effective density of the
system due to the formation of dense clusters near the pinning
centers. With an increase in the depth of the well to $w=50.0$,
the melting scenario changes from a first-order isotropic liquid -
hexatic transition to a continuous transition of the BKT type. As
a result, the Mayer-Wood loop is transformed into a kink on the
equation of state. For $w=50.0$, the pressure-density dependence
demonstrates behavior characteristic of both a first-order
transition with a Mayer-Wood loop and a continuous crossover
observed for a system with deeper pinning centers with $w=100$,
$200$. It is interesting to note that in these cases, the pressure
both in liquid and in the hexatic phase at high densities turns
out to be less than for the system without pinning or with the small well depth of the pinning centers, which is due to the above-mentioned decrease
in the average density of the system. However, in the crossover area directly, the pressure for the system with deep pinning centers proves to be higher. Our opinion is that it may be so because under gradual crossover the system in this area of densities inherits the properties of the initial liquid characterized by a greater degree of disorder than the resulting hexatic phase (and greater than in the two-phase region of first-order transition.)

The result of increased pressure in the crossover region for the
system with $w=100$ and $200$ is more pronounced ordered dense
clusters in the vicinity of the pinning centers, see Fig.
\ref{w200h}.

The cluster sizes for wells $w=50.0$, $100.0$, $200.0$ are in good
agreement with the width of the Gauss potential well in terms of
$U_{pin}$ ~ $-2$ - $-5$ ($r_{cg}=1.8$  - $1.9$), significantly
exceeding the average kinetic energy of the particles.

We summarized all our results and presented them
in Fig. \ref{phd} in the form of a phase diagram in the $w-\rho$ plane.

\section{Conclusions}

In the present paper we have studied the influence of Gaussian
random pinning on the melting behavior of 2D soft disks with $n=12$.
We introduced into the system a number of pinning centers which
attract the particles of the system via the Gauss potential with
the different depths of the wells (the width of the Gauss potential
well). We found that the influence of Gaussian wells affected the
structure of the system significantly. When the depth of the well
increases to $w=50.0$ the melting scenario of the system changes
from the BK one (a continuous BKT transition from the crystal
to the hexatic phase and a first-order phase transition from the
hexatic phase to isotropic liquid) to the BKTHNY scenario (two
continuous BKT transitions). Thus, it has been shown that it is
possible to observe various scenarios of melting in 2D
systems.

Thus, the introduction of disorder with the Gauss
potential into systems leads to an increase in the width of the hexatic phase
region and the formation of dense clusters in the vicinity of the pinning
centers that lower the average density of the system and increase the density of the solid-hexatic transition. The most intriguing property of the phase diagram of Fig. \ref{phd} is the change of the melting scenario from BK to BKTHNY. It seems that the appearance of dense clusters around the pinning centers can increase the energy of the disclination core and corresponding transformation of first-order to BKT transition \cite{ourphysusp,ourjetp,minh87,book11}, however, this issue needs further study.
The results should be interesting for materials science applications, including the formation of anisotropic fluids with different a short-range translational and a quasi-long-range orientational ($n$-fold) order and experimental studies of quenched disorder influence on the self-assembly of two-dimensional systems in the presence of controlled interactions induced by external electric \cite{jcis} or magnetic \cite{natmat} fields, as well as their combinations. In addition, it is possible to adjust interactions between particles using external fields. Taking into account natural and "man-made" substrate defects will make it possible to predict experimental conditions for detecting the boundaries of the phase diagram regions and investigating the melting scenarios of crystalline phases. Thus, understanding the behavior of particles in the presence of pinning will expand the use of such systems for technological research in the modern sciences as well as for the creation of new materials.

\section{Declaration of Competing Interest}

The authors declare that they have no known competing financial interests or personal relationships that could have appeared
to influence the work reported in this paper.

\section{Acknowledgments}

This work was carried out using computing resources of the federal
collective usage center "Complex for simulation and data
processing for mega-science facilities" at NRC "Kurchatov
Institute", http://ckp.nrcki.ru, and supercomputers at Joint
Supercomputer Center of the Russian Academy of Sciences (JSCC
RAS). The work was supported by the Russian Science Foundation
(Grant 19-12-00092, https://rscf.ru/project/19-12-00092).

\end{document}